# Using Recurrent Neural Networks To Forecasting of Forex

V.V.Kondratenko[1] and Yu. A Kuperin[2]

[1] *Division of Computational Physics, Department of Physics, St.Petersburg State University*

[2] *Laboratory of Complex Systems Theory, Department of Physics, St.Petersburg State University*
*E-mail: kuperin@JK1454.spb.edu*

## Abstract

This paper reports empirical evidence that a neural networks model is applicable to the statistically reliable prediction of foreign exchange rates. Time series data and technical indicators such as moving average, are fed to neural nets to capture the underlying "rules" of the movement in currency exchange rates. The trained recurrent neural networks forecast the exchange rates between American Dollar and four other major currencies, Japanese Yen, Swiss Frank, British Pound and EURO. Various statistical estimates of forecast quality have been carried out. Obtained results show, that neural networks are able to give forecast with coefficient of multiple determination not worse then 0.65. Linear and nonlinear statistical data preprocessing, such as Kolmogorov-Smirnov test and Hurst exponents for each currency were calculated and analyzed.

## 1. Introduction

Forex is the largest and most liquid of the financial markets, with an approximately $1 trillion traded every day. It leads to the serious interest to this sector of finance and makes clear that for various reasons any trader on Forex wish to have an accurate forecast of exchange rate. Most of traders use in old fashion manner such traditional method of forecast as technical analysis with the combination of fundamental one. In this paper we develop neural network approach to analysis and forecasting of financial time series based not only on neural networks technology but also on a paradigm of complex systems theory and its applicability to analysis of various financial markets (Mantegna et al., 2000; Peters, 1996) and, in particular, to Forex. While choosing the architecture of neural network and strategy of forecasting we carried out data preprocessing on the basis of some methods of ordinary statistical analysis and complex systems theory: R/S-analysis, methods of nonlinear and chaotic dynamics (Mantegna et al., 2000; Peters, 1996). In the present paper we do not describe all of them. We present here only the results of the Kolmogorov-Smirnov test and results of R/S-analysis. However we stress that the preliminary analysis has allowed to optimize parameters of neural network, to determine horizon of predictability and to carry out comparison of forecasting quality of different currencies.

Below we give some remarks relating to advantages of neural networks technology over traditional methods and compare our approach with the methods of other authors

First, neural networks analysis does not presume any limitations on type of input information as technical analysis does. It could be as indicators of time series, as information about behavior of another financial instruments. It is not without foundation, that neural networks are used exactly by institutional investors (pension funds for example), that deal with big portfolios and for whom correlations between different markets are essential.

Second, in contrast to technical analysis, which is based on common recommendations, neural networks are capable to find optimal, for given financial instrument, indicators and build optimal, for given time series, forecasting strategy.

Let us remind that in present study we forecasted the exchange rates of only selected currencies on Forex market. As currencies to deal with, we chose British Pound, Swiss Frank, EURO and Japanese Yen. The following motivates this choice: practically all the main volumes of operations on Forex are made with this currencies.

Let us note that there were published a lot of papers, where similar problems have been studied. (Jingtao Yao et al., 2000; Castiglione, 2001; Kuperin et al., 2001; Lee et al., 1997; Tino et al., 2001; McCluskey, 1993). Let us briefly look round the results some of them.

In (Castiglione, 2001) there were studied the problem of sign of price increments forecasting. As an analyzed data were taken such indices as S&P500, Nasdaq100 and Dow Jones. Has been taken multilayer perceptrons of different configurations, with different number of hidden neurons. As a result there has been shown a possibility of forecasting of the sign of price increments with probability of slightly higher than 50%, i.e. a little bit better then just coin tossing. We suppose, that such kind of results is irrelevant from the practical point of view and has an academic interest.

In (McCluskey, 1993) has been studied the problem of forecasting the price increments of S&P500. As the analyzed data has been taken historical S&P500 data for the period from 1928 to 1993. There have been built and trained a lot of neural networks of different configuration. There has been estimated the profitability of using the neural networks in question. In the result surprisingly almost all networks were able to give a profitable forecast. Any statistical estimates of forecast quality were absent. It is clear that statistical estimates of forecast quality are extremely important, since profitable forecast might be just accidental.

In (Tino et al., 2001) there has been forecasted the sign of price increments. As the analyzed data there have been taken DAX and FTSE100 for some time period. The Elman recurrent neural network has been chosen. To the input there were fed binary signals corresponding to the sign of price increments. As an estimate of forecast quality, the profitability was chosen as in above paper. In the result the authors made a conclusion, that neural networks are not capable to give better results than more simple models, such as Markov models for example.

In (Jingtao Yao et al., 2000) there were carried out a neural network forecast of currencies exchange rates on Forex. As the analyzed data there were chosen weekly data of AUD, CHF, DEM, GBP and JPY. Two neural models in this study have been used:
1. Multilayer perceptron, to inputs of which there were fed values of exchange rates with some time lag.
2. Multilayer perceptron, to inputs of which there were fed values of moving average of exchange rates with different time windows (from 5 to 120 ticks).
In the result there were performed statistical estimates of forecast quality, which showed for the last model, that neural network is able to give a good result. At the same time we have to emphasize, that forecast of the exchange rate by itself is of no practical value. Besides of that, the forecast of weekly data presumes, that trader, who uses this forecast will trade once a week, which is irrelevant from the practical point of view.

We have also to mention, that forecasting of future values of underlying time series gives, besides the direct profit, the opportunity of calculating of some interesting quantities, such as price of derivative or probability of adverse mode, which is essential when one estimate the risk of investment. (Castiglione, 2001).

In the reminder of the paper we describe the structure of financial data we used, the data preprocessing that we carried out, i.e. normalization, calculation of Hurst exponent, Kolmogorov-Smirnov test. Next we describe the neural network architecture we used and statistical estimates of the results. Finally we present the results of time series forecasting and discussion. Last section concludes.

# 2. Time series forecasting

## 2.1 Structure of Data

In this section we remind the well-known features of economic and financial time series, which should be taken into account in training of neural networks and appropriate forecasting. What is follows are the major features:

*Noise*: Macroeconomic time series are intrinsically very noisy and generally have poor signal to noise ratios. The noise is due both to the many unobserved variables in the economy and to the survey techniques used to collect data for those variables that are measured. The noise distributions are typically heavy tailed and include outliers. The combination of short data series and significant noise levels makes controlling model variance, model complexity, and the bias / variance tradeoff important issues (Geman, Bienenstock and Doursat, 1992). One measure of complexity for nonlinear models is $P_{eff}$, the effective number of parameters (Moody, 1992; Moody, 1994). $P_{eff}$ can be controlled to balance bias and variance by using regularization and model selection techniques.

*Nonstationarity*: Due to the evolution of the world economies over time, macroeconomic time series are intrinsically nonstationary. To confound matters, the definitions of many macroeconomic time series are changed periodically as are the techniques employed in measuring them. Moreover, estimates of key time series are periodically revised retroactively as better data are collected or definitions are changed. Not only do the underlying dynamics of the economy change with time, but the noise distributions for the measured series vary with time also. In many cases, such nonstationarity shortens the useable length of the data series, since training on older data will induce biases in predictions. The combination of noise and nonstationarity gives rise to a noise / nonstationarity tradeoff (Moody, 1994), where using a short training window results in too much model variance or estimation error due to noise in limited training data, while using a long training window results in too much model bias or approximation error due to nonstationarity.

*Nonlinearity*: Traditional macroeconomic time series models are linear (Granger and Newbold, 1986; Hamilton, 1994). However, recent works by some authors have suggested that nonlinearities can improve macroeconomic forecasting models in some cases (Granger and Terasvirta, 1993; Moody et al., 1993; Natter, Haefke, Soni and Otruba, 1994; Swanson and White, 1995). (See table 1 and figures 2 and 3.) Based upon our own experience, the degree of nonlinearity captured by neural network models of macroeconomic time series tends to be mild (Moody et al., 1993; Levin, Leen and Moody, 1994; Rehfuss, 1994; Utans, Moody, Rehfuss and Siegelmann, 1995; Moody, Rehfuss and Saffell, 1996; Wu and Moody, 1996). Due to the high noise levels and limited data, simpler models are favored. This makes reliable estimation of nonlinearities more difficult.

## 2.2 Data preprocessing

As it was mentioned above, in this paper there have been analyzed exchange rates of four main currencies on Forex: British pound, Swiss frank, EURO and Japanese Yen in the period from 22.07.1998 to 02.09.2001. Data was taken from www.gelium.net and www.forexite.com.

There are diagrams of exchange rates of currencies in question below for entire period and for shorter period. It could be seen that in spite of apparent smoothness graphs are rather heavily jagged.

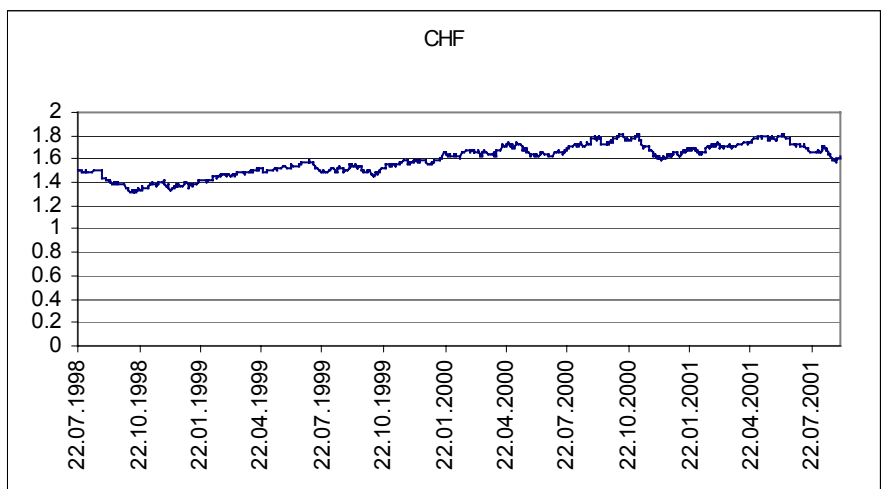

Figure 1a. Daily exchange rate of CHF for entire period 22/07/1998 – 02/09/2001. Along the coordinate axes are given logarithmic returns (ordinate axis) and time (abscissa axis), the definition see below

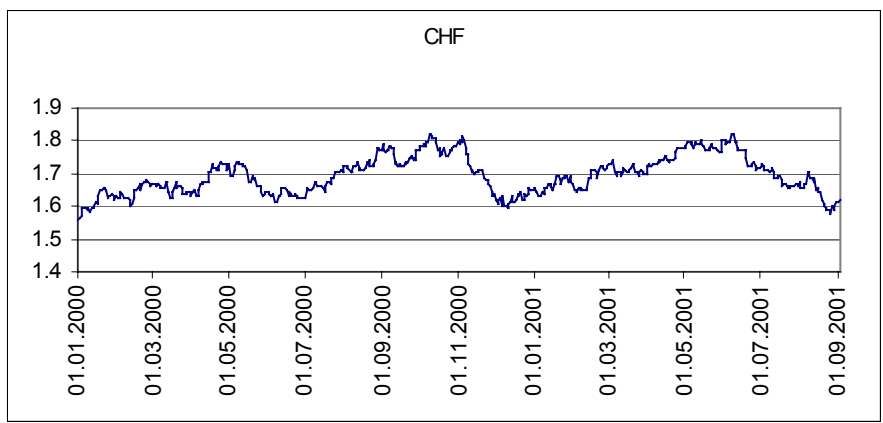

Figure 1b. Daily exchange rate of CHF for shorter period 01/01/2000–01/09/2001. Along the coordinate axes are given logarithmic returns (ordinate axis) and time (abscissa axis), the definition see below

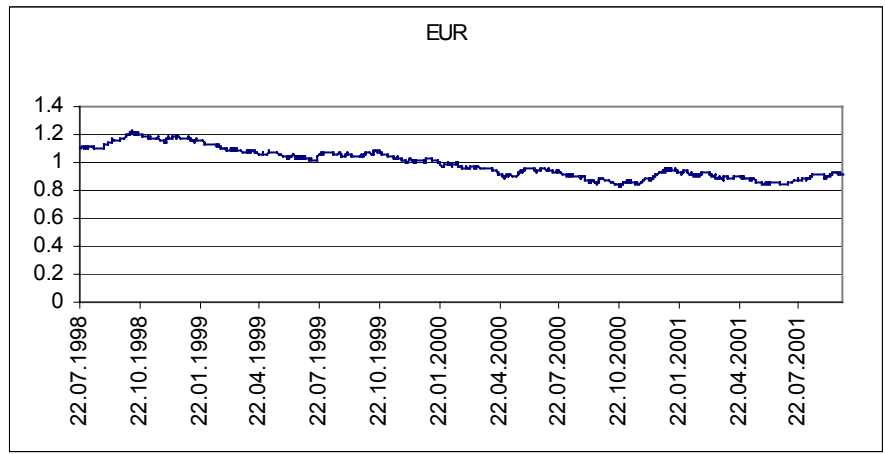

Figure 2a. Daily exchange rate of EUR for entire period 22/07/1998 – 02/09/2001. Along the coordinate axes are given logarithmic returns (ordinate axis) and time (abscissa axis), the definition see below.

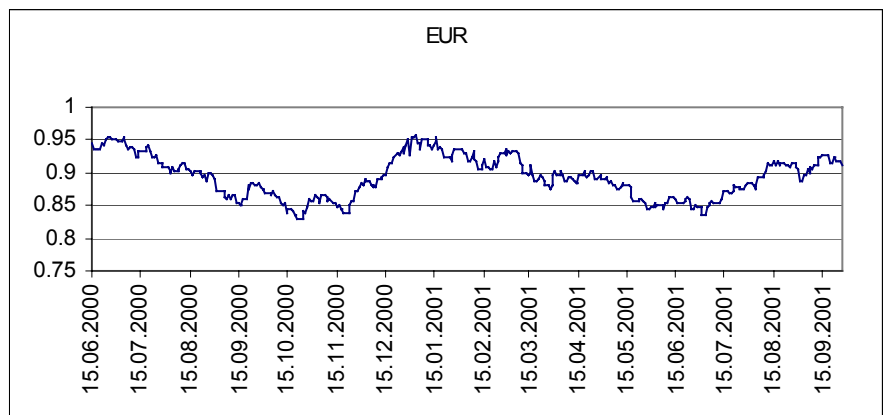

Figure 2b. Daily exchange rate of EUR for shorter period 15/06/2000–15/09/2001. Along the coordinate axes are given logarithmic returns (ordinate axis) and time (abscissa axis), the definition see below.

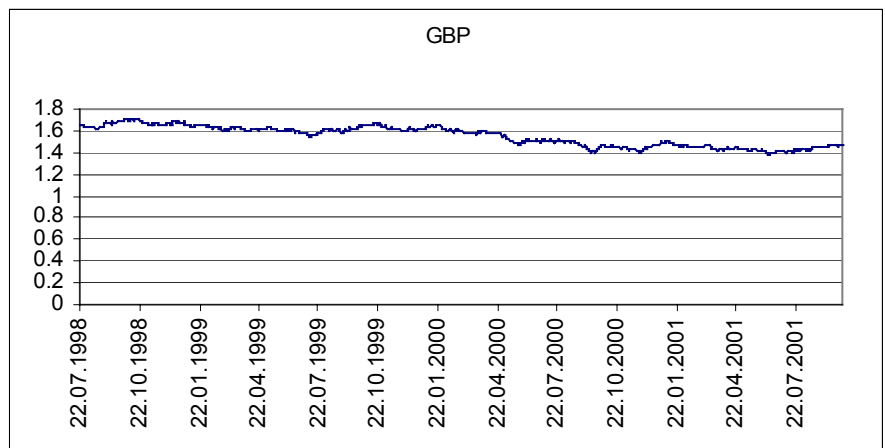

Figure 3a. Daily exchange rate of GBP for entire period 22/07/1998 – 02/09/2001. Along the coordinate axes are given logarithmic returns (ordinate axis) and time (abscissa axis), the definition see below.

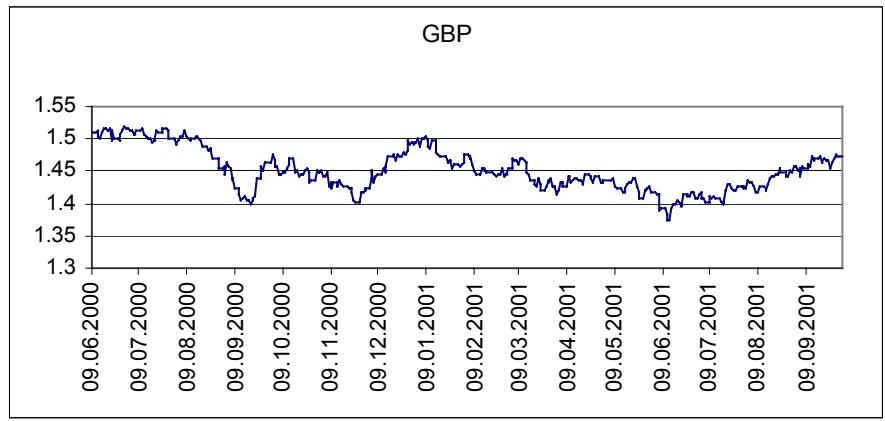

Figure 3b. Daily exchange rate of GBP for shorter period 09/06/2000–09/09/2001. Along the coordinate axes are given logarithmic returns (ordinate axis) and time (abscissa axis), the definition see below.

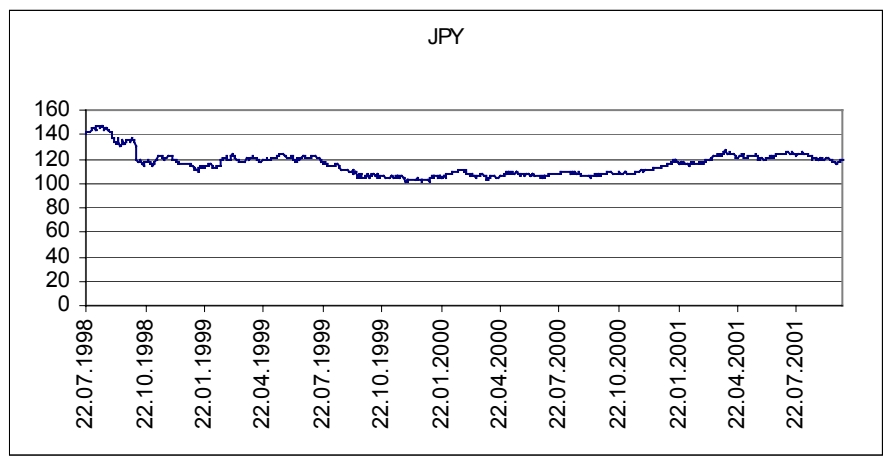

Figure 4a. Daily exchange rate of JPY for entire period 22/07/1998 – 02/09/2001. Along the coordinate axes are given logarithmic returns (ordinate axis) and time (abscissa axis), the definition see below.

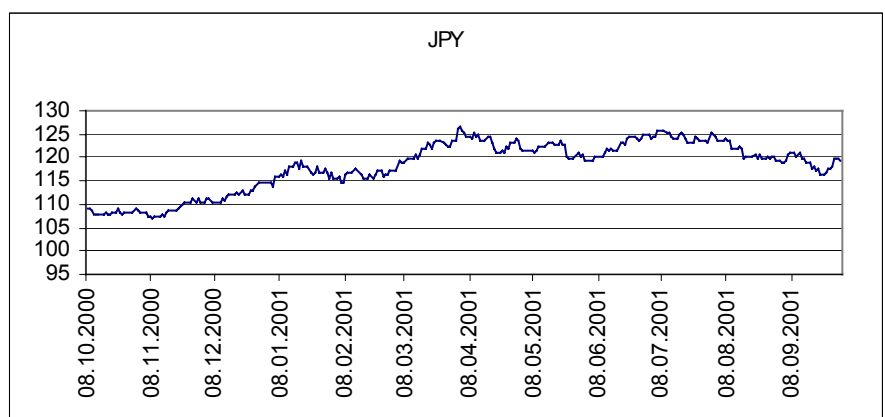

Figure 4b. Daily exchange rate of JPY for shorter period 08/10/2000–08/09/2001. Along the coordinate axes are given logarithmic returns (ordinate axis) and time (abscissa axis), the definition see below.

Financial data has particular property: the higher is the frequency of data collection the worse is signal/noise ratio. Regard to this, weekly data are usually better forecasted, than daily data. On the other hand the more accurate is the forecast the higher is its practical value. In compliance with these principles we chose daily closing prices as input data.

Usually traders are not really interested in forecast of exchange rate itself. Much more of interest is the increment of rate or the sign of increment forecast. Moreover, there is strong correlation between current and previous values of price – most probable value of price in next moment equals to its previous value. Meanwhile, in order to improve the quality of forecast it is necessary to make inputs statistically independent, i.e. to remove correlations.

Because of this in our study we analyzed not the exchange rates but its logarithmic returns:

$$r_n = \ln(\frac{S_n}{S_{n-1}}) \qquad (1)$$

It is clear, that results of neural network analysis should not depend on the units of measurement. It means that all the input and output values should be reduced to a unit scale, in other words to be normalize. Moreover, in order to improve the learning rate it is effective to carry out additional data preprocessing that smoothes the data distribution before learning phase.

In the present study normalization of data was performed with the use of logistic function – the activation function of neurons. Transformation was performed corresponding to following formula:

$$\tilde{x}_i = \frac{1}{1 + \exp(-\frac{x_i - \bar{x}}{\sigma})} \qquad (2),$$

where: $\bar{x}$ is the mean, $\sigma$ is the standard deviation.

This transformation normalizes main volume of data and guarantees that $\tilde{x}_i \in [0,1]$. Moreover it increases the entropy of data, since after transformation data fills the interval [0,1] more even.

The most well known but little believed hypotheses are Random Walk Hypothesis and Efficient Market Hypothesis. The Random Walk Hypothesis states that the market prices wander in a purely random and unpredictable way. The Efficient Market Hypothesis roughly speaking states that markets fully reflect all of the available information and prices are adjusted fully and immediately once new information become available. In the actual market, some people do react immediately after they have received the information while other people wait for the confirmation of information. The waiting people do not react until a trend is clearly established. All together (in week form or in stricter form) it means that in frames of Efficient Market Hypothesis is nothing to do with any type of forecasting and, in particular, by means of neural networks. However, our study is based on the recent empirical evidence that in actual markets, the residual inefficiencies (caused by different reasons) are always present. In general we are not going to discuss these reasons here, however we shall use some of tools from linear and nonlinear statistics in order to analyze the deviation of data in question from efficiency. That is the standard Kolmogorov-Smirnov test and R/S analysis (or Hurst exponents).

The Hurst Exponent is in particular a measure of the bias in the fractional Brownian motion (Peters 1996, Peters 1994, Kantz 1997). In general (modulo any probability distribution) the method could be used in economic and financial market time series to know whether these series are biased random walk, which indicates in some sense the possibility of forecasting.

The Hurst Exponent H describes the probability that two consecutive events are likely to occur (for more details see, e.g. Peters 1996). There are three distinct classifications for the Hurst exponent: 1) H=0.5, 2) $0 \leq H < 0.5$, and 3) $0.5 < H < 1.0$. The type of series described by H=0.5 is a pure random, consisting of uncorrelated events. However, value H=0.5 cannot be used as an evidence of a Gaussian random walk. It only proves that there is no evidence of a long memory effect. A value of H different from 0.5 denotes that the observations are correlated. When $0 \leq H$

≤ 0.5, the system is characterized by antipersistent or ergodic time series with frequent reversals and high volatility. At the same time despite the prevalence of the mean reversal concept in economic and finance literature, only few antipersistent series have yet been found. For third case (0.5 ≤ H ≤ 1), H describes a persistent or trend reinforcing series, which is characterized by long memory effects. The strength of the bias depends on how far H is above 0.5. However, even in the case that the Hurst process describes a biased random walk, the bias can change abruptly either in direction or magnitude. Therefore, only the average cycle length of observed data can be estimated.

It is useful before neural network forecasting to check if our time series are persistent or antipersisnet and whether returns have Gaussian distribution. This can give some information on the predictability of time series in question. If time series have H≈0.5 and returns have Gaussian distribution then we can presume that these time series hardly could be accurately forecasted. On the other hand, the closer H is to 1 and the more the distribution of returns differs form Gaussian distribution, the higher is the probability that these time series could be forecasted with high quality.

So to check the predictability of the analyzed time series there has been performed statistical analysis of data. For each of the currencies there were computed first four moments, i.e. mean, variance, skewness and kurtosis. Also there was performed the Kolmogorov-Smirnov test and have been computed Hurst exponents. Results are given below.

Table 1.1. *Statistical data for currencies*.

|     | Mean     | Maximum  | Minimum  | Variance | St. Deviation | Skewness  | Kurtosis |
|-----|----------|----------|----------|----------|---------------|-----------|----------|
| CHF | -8,3E-05 | -0,03798 | 0,035216 | 4,92E-05 | 0,007017      | 0,238086  | 2,291049 |
| EUR | 0,000246 | -0,02335 | 0,023201 | 3,2E-05  | 0,005661      | -0,18691  | 2,138603 |
| GBP | -1,7E-05 | -0,02245 | 0,024826 | 2,42E-05 | 0,004917      | 0,009823  | 1,655181 |
| JPY | 4,13E-05 | -0,07811 | 0,035251 | 6,01E-05 | 0,007755      | -0,94684  | 7,928037 |

Table 1.2 *The Kolmogorov-Smirnov test results*.

| CHF | 0.05871  |
|-----|----------|
| EUR | 0.010531 |
| GBP | 0.04967  |
| JPY | 0.06328  |

Table 1.3 *The values of Hurst Exponents*

| CHF | 0.566 ± 0.002 |
|-----|---------------|
| EUR | 0.570 ± 0.002 |
| GBP | 0.523 ± 0.004 |
| JPY | 0.557 ± 0.002 |

As it is shown in Tables 1.1-1.3, the values of Hurst exponents for the logarithmic returns of daily exchange rates data is higher than 0.5 for all the observed time series. This shows that all the studied currencies are persistent. The highest value is 0.570 for the exchange rate of EUR/USD. In other words, it is the least efficient currency amongst the four studied ones. The results of the Kolmogorov-Smirnov test show that not any returns of observed time series have Gaussian distribution. This means that not any time series are described by Brownian motion. The closest distribution to Gaussian is the distribution of EUR logarithmic returns. It means, that EUR returns are weaker correlated than the others currencies time series returns and probably EUR will have the worst forecast.

The results obtained in this part of the study imply that the all time series are nor antipersistent neither Brownian motion. Thus neural network forecasting of the exchange rates is possible and should has a good quality.

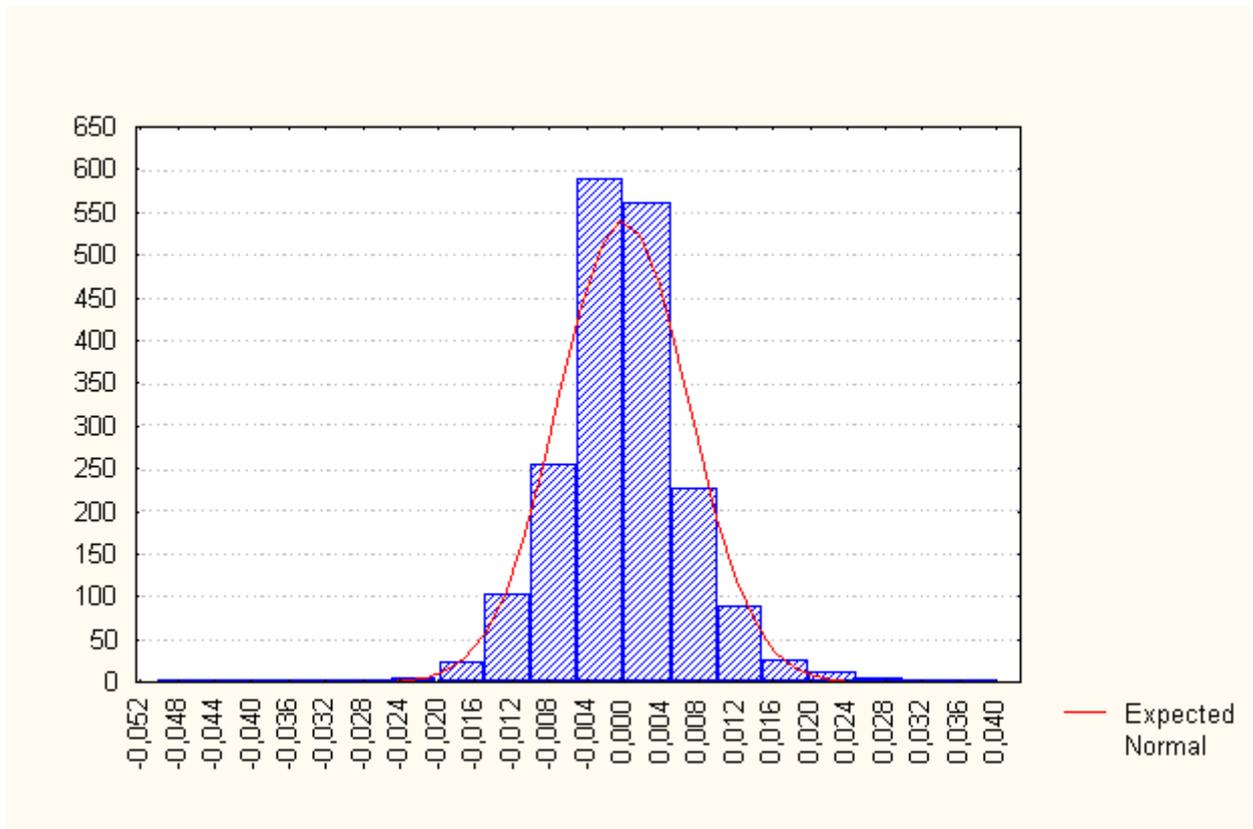

Figure 5. Distribution of CHF. Here abscissa axis is the upper boundaries of fragmentation of the logarithmic returns and ordinate axis is the number of observations of return's value in each fragmentation interval.

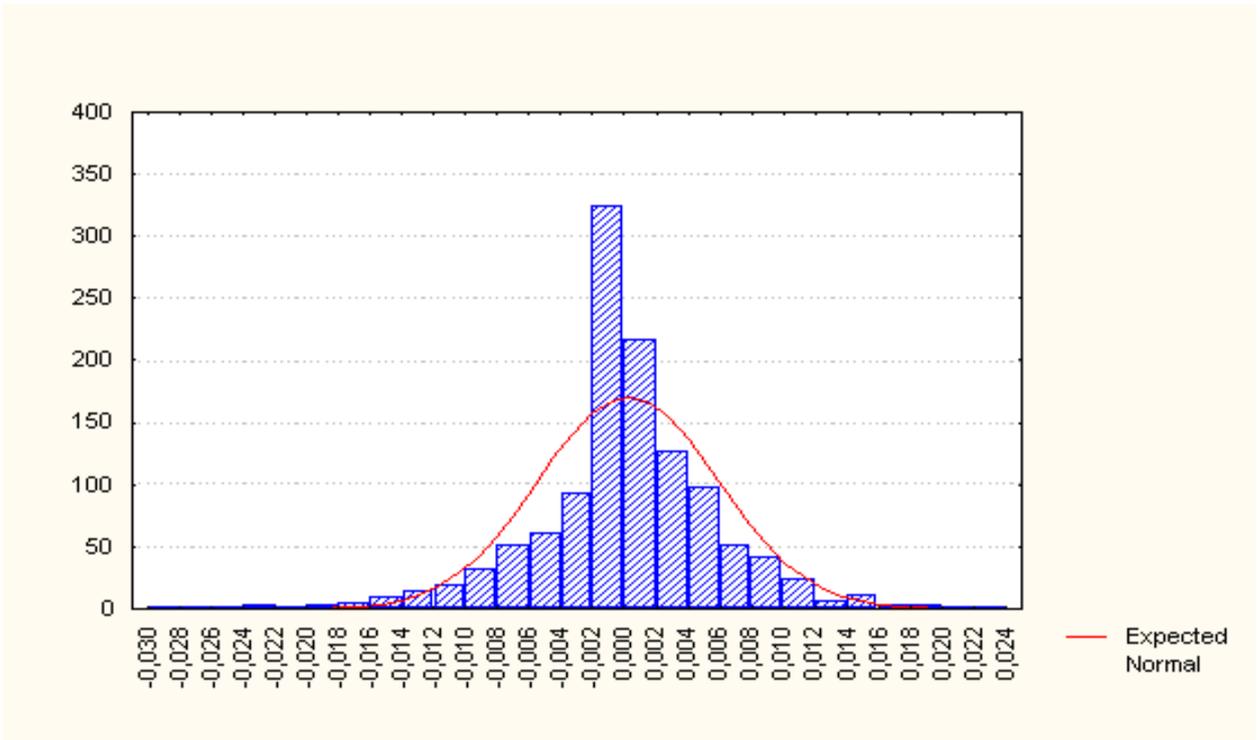

Figure 6. Distribution of EUR. Here abscissa axis is the upper boundaries of fragmentation of the logarithmic returns and ordinate axis is the number of observations of return's value in each fragmentation interval.

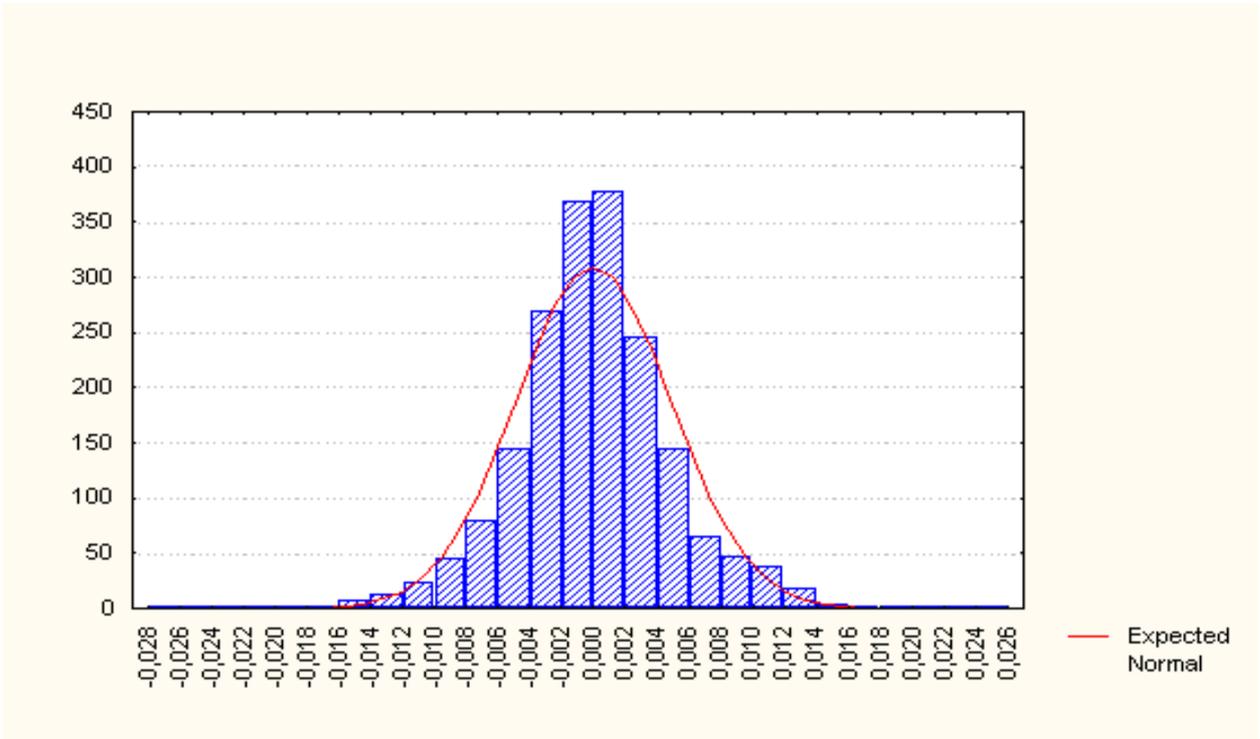

Figure 7. Distribution of GBP. Here abscissa axis is the upper boundaries of fragmentation of the logarithmic returns and ordinate axis is the number of observations of return's value in each fragmentation interval.

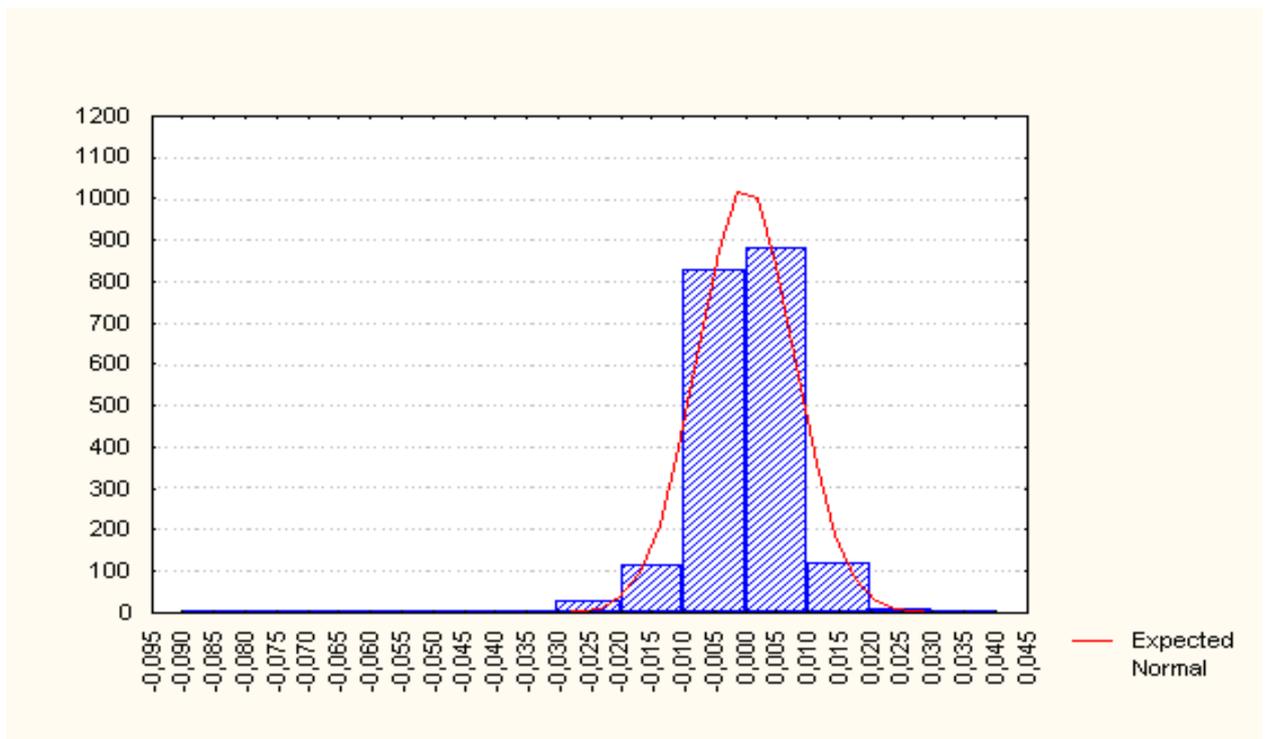

Figure 8.Distribution of JPY. Here abscissa axis is the upper boundaries of fragmentation of the logarithmic returns and ordinate axis is the number of observations of return's value in each fragmentation interval.

## 1.3 Choosing the of neural network architecture

The presence of recurrent feedback in neural network seems to be a positive factor in forecasting of financial time series. (Castiglione, 2001; McCluskey, 1993)  This is apparently because the recurrent neural network has a "deeper memory" than feedforward neural network.

In our study the Elman-Jordan neural network has been used. This class of neural networks with learning by back propagation method was successfully used for prediction of financial markets since it is the recurrent neural net which learns the rules in the time series, which is necessary when works with ones. The disadvantage of this class of neural nets is the long learning time. (Kuperin et al., 2001).

The network we built had two inputs and one output. To one input there were fed daily returns, to another one there were fed daily returns, smoothed by moving average with the window equal to 5 observations. Moving average in input was supposed to smooth the noisy time series of returns. As the output we chose the value of moving average of returns with the window equal to 5 observations, shifted to one day forward, because it corresponds to trading purposes better than returns because of more accurate forecast and such fact that if trader would try to trade on every market movement it would lead to very expensive transaction costs., It is partially explain why it is reasonable to forecast the moving average. Thus, the network was supposed to forecast the one-day ahead value of moving average of returns with the window equal to 5 observations. The number of hidden neurons was chosen equal to 100. .

Neurons of input layer had linear activation functions. Neurons of hidden and output layers had logistic activation function. This configuration was determined experimentally as giving the best results.

Momentum and learning rate were chosen equal to 0.003. The net terminated its learning if during 20000 epochs it was not capable to improve its results. Training, test and production sets were chosen as following: first 1000 observations were used as the training set, next 200 observations were used as the test set, and last 100 observations were used as the production set.

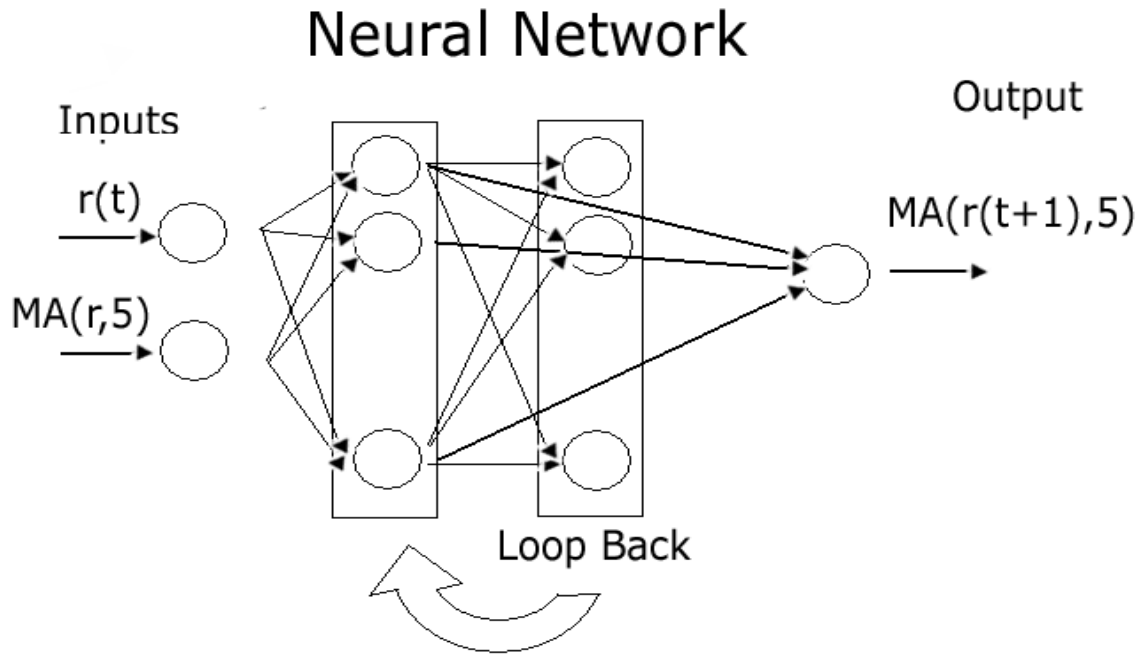

Figure 9. Schematic illustration of the Elman-Jordan network architecture. Here r(t) is logarithmic return, MA(r,5) is the moving average with window equal to 5 of logarithmic returns and *Loop Back* denotes the interlayer feedback.

## 2.4 Statistical estimation of forecasting results

In the present section we briefly describe some indices we used for statistical estimations of the neural network forecasting quality. This part of the study is quite important both for the proper choice of neural network architecture and for statistical relevance of obtained results. The specific values of obtained indices are discussed below.

In particular, as performance criteria there were chosen the following indices:

***$R^2$*** The coefficient of multiple determination is a statistical indicator usually applied to multiple regression analysis. It compares the accuracy of the model to the accuracy of a trivial benchmark model wherein the prediction is just the mean of all of the samples. A perfect fit would result in an R squared value of 1, a very good fit near 1, and a very poor fit less than 0. If the neural model predictions are worse than one could predict by just using the mean of the sample case outputs, the R squared value will be less than 0. The formula used for $R^2$ is given by

$$R^2 = 1 - \frac{\sum_i (y_i - \hat{y})^2}{\sum_i (y_i - \bar{y})^2} \qquad (3)$$

where: $y_i$ is the actual value, $\hat{y}$ is the predicted values of $y_i$, $\bar{y}$ is the mean of the $y_i$ values

***r squared*** This is the square of the correlation coefficient, described below.

***Mean Squared Error*** This is the mean over all patterns of the square of the actual value minus the predicted value, i.e., the mean of (actual minus predicted) squared.

*Min Absolute Error* This is the minimum of actual minus predicted values over all patterns.

*Max Absolute Error* This is the maximum of actual minus predicted values over all patterns.

*Correlation Coefficient r* (Pearson's Linear Correlation Coefficient) This is a statistical measure of the strength of the correlation between the actual and predicted outputs. The r coefficient can range from -1 to +1.

*Percent within 5%, 10%, 20% and 30% and over 30%* It is the percent of network answers that are within the specified percentage of the actual answers used to train the network. If the actual answer is 0, the percent cannot be computed and that pattern is not included in a percentage group. For that reason and rounding, the total computed percentages may not add up to 100%.

*Ability of correct prediction for the sign of increments* This is the number of correctly predicted signs of increments, divided by the whole number of patterns.

Most representative statistical estimate of forecast quality among of all mentioned above is $R^2$. Other representative indices are: r squared, mean absolute error and maximum absolute error. Mean absolute error and maximum absolute error show the accuracy of forecast, r squared shows the linear correlation, but tells nothing about accuracy. $R^2$ gives information both on accuracy and correlation. It can be shown that

$$\bar{\varepsilon} \leq \sqrt{(1-R^2)}\,\sigma \quad (4)$$

where: $\bar{\varepsilon}$ is the mean absolute error, $\sigma$ is the standard deviation.
For example, if we have $R^2 = 0.8$, so we can estimate mean absolute error as $\bar{\varepsilon} \leq 0.447\sigma$.

## 3. Results and discussion

In the present study there were built and trained four neural networks, one for each currency. In Appendix we present the Tables with the statistical results of forecast quality.

Below we give also the examples of graphs, which visualize the quality of forecast, where ordinate axis is the value of logarithmic returns and abscissa axis is the time.

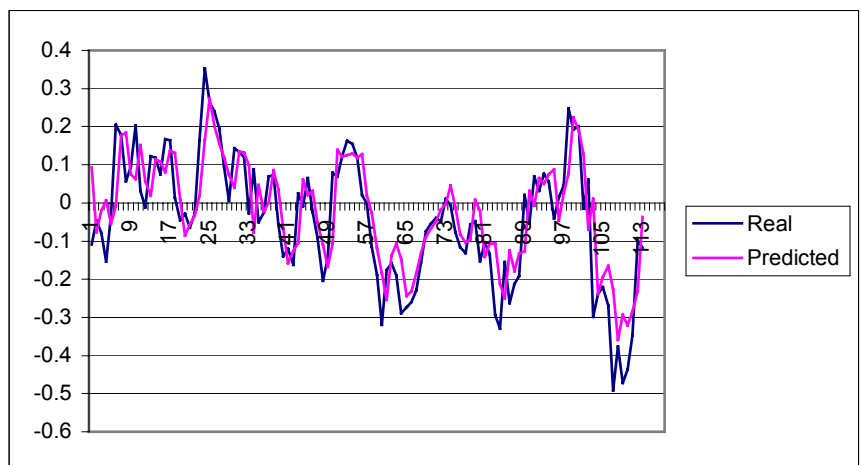

Figure 10. Actual and predicted values of CHF for the period from 18/04/2001 to 01/10/2001.

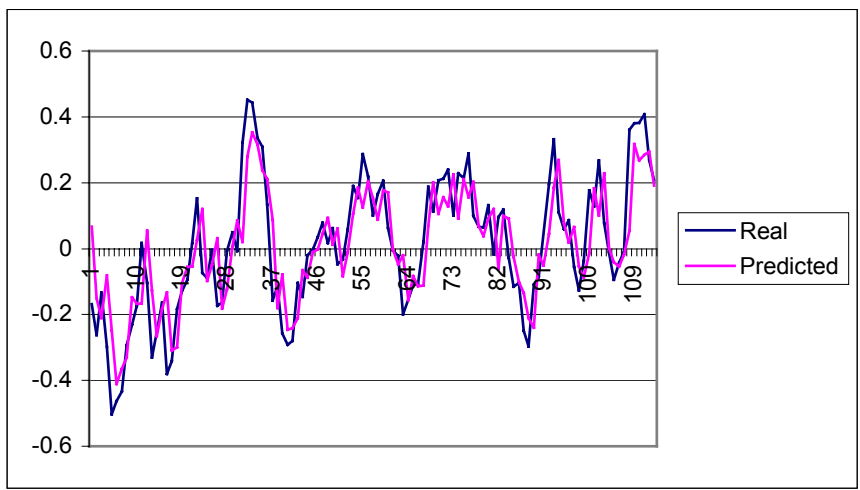

Figure 11. Actual and predicted values of EUR for the period from 12/08/2001 to 04/05/2002.

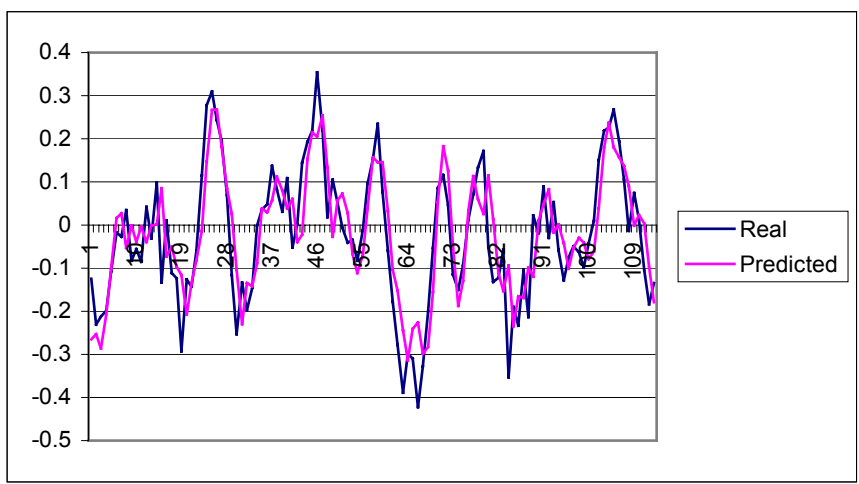

Figure 12. Actual and predicted values of GBP for the period from 18/04/2001 to 01/10/2001.

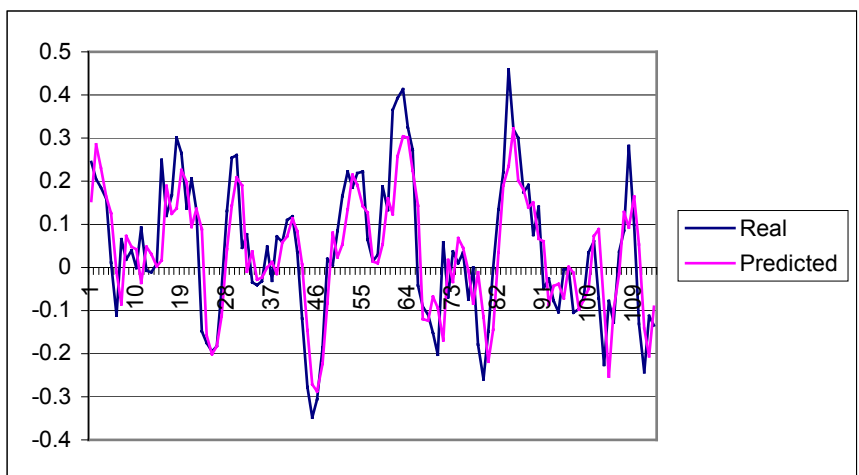

Figure 13. Actual and predicted values of JPY for the period from 23/04/2001 to 01/10/2001.

We also made some attempts to improve the achieved results in the following direction. Since the amount of data in use was limited by available data on EURO (exchange rate of EURO appeared in 1998), we decided to take more data in spite of its different amount for different currencies. Thus, the amount of data in use was chosen equal to 2200 daily ticks. Moreover, we decided to change the configuration of inputs and to add some technical indicators to the input of the neural net. We have chosen the Relative Strength Index (RSI) and Williams Percent Range (%R) as such inputs. In other words there were built several neural networks with different configuration of inputs. Below we give more detailed description of these models. (See Appendix for the appropriate results and statistical estimation). Namely, we undertaken the following numerical experiment:

1. We have added to the input of neural network such technical indicators as RSI and %R. We expected to improve the quality of forecast due to presence of leading technical indicators in the network input. Since the number of inputs increased, we also increased the number of hidden neurons up to 140.

2. As the next step we have decided to feed only technical indicators in input of the network. Moving average of returns and returns themselves were removed. We supposed to check the prevalent belief that all the necessary information contains in RSI and % R time series and so the excess information could only worsen the quality of forecast.

3. Since it turned out that the results, shown by previous model were essentially worse than the others, we decided to feed to the network input the technical indicators and moving average of returns, removed at the same time returns themselves from the input. We supposed that presence of raw data in the input could worsen the quality of forecast.

4. Since the best results have been shown by the network, which had two indicators, i.e., moving average and returns in input, we have decided to take this model as the basic one and only slightly modify it. We replaced moving average by exponential moving average, since it approximates time series more accurate and contains information on longer time period.

5. Finally, we have used the embedding method. Namely, we fed four series values of returns and supposed to obtain the next day returns from the output.

## 4.Conclusions

Let us summarize the obtained results as follows.
The statistically best results have been obtained by the model, where logarithmic returns and moving average were taken as inputs. Comparing the results, obtained from this model, to others ones it should be noted that:
1. *Model with two added technical indicators:* in general, the forecast quality became worse than the initial model ($R^2$ is increased on CHF only, on the others currencies it decreased). It turned out that the probability of correctly predicted increments sign is also lower. Apparently, the information that contains in indicators is excess and makes the problem of forecast more complicated. We could increase, of course, the number of hidden neurons, but it would lead to unwarranted increasing of the learning time.
2. *Model with only RSI and %R in input:* statistically the forecast quality fell down even more then in other models we studied. This indicates, that the mentioned above prevalent belief about information content of technical indicators is irrelevant to the neural network forecasting.
3. *Model with RSI, %R and moving average in input:* comparing to previous models the forecast quality has been improved, but it still remained not the best. It looks like the presence of moving average in input of the network is the positive factor at least due to the smoothing of the data.
4. *Model with indicators, returns and exponential moving average:* results, obtained from this model are analogous to results obtained by basic model. Apparently, there is we have found no essential difference between ordinary moving average and exponential moving average for the problem in question.
5. *Embedding method:* the particular property of this model is that the results of forecasting of different currencies are much more analogous, than in other models. Nevertheless the forecast quality is worse than in basic model. Apparently, it is possible to obtain better results from this model by increasing the number of inputs and hidden neurons, but it would lead again to unjustified long learning time.

It should be noted, that CHF and JPY are usually better predicted than GBP and EURO. Is seemingly means that regularities, that contain CHF and JPY time series are more complicated than regularities, containing in GBP and EURO time series. In EURO case comparatively lower quality of forecast could be explained by insufficient amount of data.

Thus, in this study neural net forecast of four main currencies on Forex market was carried out. Finally let us conclude that statistical estimates of Forex forecast indicates, that neural network can predict the increments sign with relatively high probability – approximately 80%, which is quite sufficient for practical use. The coefficient of multiple determination $R^2$ in the best our model varied from 0.65 to 0.69. . We believe that such quality of forecast is enough for building an automatic profitable trade strategy. This point of view has been confirmed in (Kuperin et al., 2001) where a profitable trade strategy for GBP, based on forecast with similar quality has been constructed and tested.

# References

1. R.N.Mantegna, H.E.Stenley: An Introduction to Econophysics. Correlations and Complexity in Finance. Cambridge University Press (2000)
2. E.E.Peters: Chaos and Order in Capital Markets. John Wiley&Sons (1996)
3. E.E.Peters: Fractal Market Analysis. John Wiley&Sons (1994)
4. H.Kanz, T.Schreiber: Nonlinear Time Series Analysis. Cambridge University Press (1997)
5. A.A.Ezhov, S.A. Shumsky. Neurocomputing and its Applications in Economics and Business. Moscow.: MEPHI, 1998.
6. Jingtao Yao, Chew Lim Tan. A Case Study on Using Neural Networks to Perform Technical Forecasting of Forex. Neurocomputing, Vol. 34, No. 1-4, pp79-98 (2000).
7. Filippo Castiglione. Forecasting Price Increments Using an Artificial Neural Network. in *Complex Dynamics in Economics*, a Special Issue of Advances in Complex Systems, 3(1), 45-56, Hermes-Oxford (2001)
8. Kuperin, Yuri A., Dmitrieva Ludmila, A., Soroka Irina V. Neural Networks in financial markets dynamical studies: Working paper series of the Center for Management and Institutional Studies. School of Management. Saint-Petersburg State University. No 2001-12
9. C.Lee Giles, Steve Lawrence, Ah Chung Tsoi. Noisy Time Series Prediction using a Recurrent Neural Network and Grammar Inference. Proceedings of IEEEE/IAFE Conference on computational intelligence for financial engineering IEEE, Piscantaway. New York, (1997). P. 253- 259.
10. John Moody. Economic Forecasting: Challenges and Neural Network Solutions. Proceedings of the International Symposium on Artificial Neural Networks. Hsinchu, Taiwan, (1995)
11. Peter C. McCluskey. Feedforward and Recurrent Neural Networks and Genetic Programs for Stock Market and Time Series Forecasting. Master's thesis, Brown University, (1993).
12. Peter Tino, Christian Schittenkopf, Georg Dorffner. Financial Volatility Trading using Recurrent Neural Networks. Pattern Analysis and Application, (2001)
13. Moody, J. The effective number of parameters: an analysis of generalization and regularization in nonlinear learning systems, in J. E. (1992)
14. Moody, J., Challenges of Economic Forecasting: Noise, Nonstationarity, and Nonlinearity, Invited talk presented at Machines that Learn, Snowbird Utah, April (1994a).
15. Moody, J., Prediction risk and neural network architecture selection, in V. Cherkassky, J. Friedman and H. Wechsler, eds, `From Statistics to Neural Networks: Theory and Pattern Recognition Applications', Springer-Verlag. (1994b)
16. Moody, J., Levin, A. and Rehfuss, S., `Predicting the U.S. index of industrial production', Neural Network World 3(6), 791--794. Special Issue: Proceedings of Parallel Applications in Statistics and Economics '93. (1993)
17. Natter, M., Haefke, C., Soni, T. and Otruba, H., Macroeconomic forecasting using neural networks, in `Neural Networks in the Capital Markets 1994'. Pi, H. and Peterson, C. (1994), `Finding the embedding dimension and variable dependencies in time series', Neural Computation pp. 509--520. (1994)
18. Swanson, N. and White, H., A model selection approach to real-time macroeconomic forecasting using linear models and artificial neural networks, Discussion paper, Department of Economics, Pennsylvania State University. (1995)
19. Rehfuss, S., Macroeconomic forecasting with neural networks. Unpublished simulations. (1994)
20. Utans, J., Moody, J., Rehfuss, S. and Siegelmann, H., Selecting input variables via sensitivity analysis: Application to predicting the U.S. business cycle, in `Proceedings of Computational Intelligence in Financial Engineering'. (1995)
21. Wu, L. and Moody, J., `A smoothing regularizer for feedforward and recurrent networks', Neural Computation 8(2). (1996)

# Appendix

Table 2. Statistics of forecast quality for main currencies of Forex (the basic model)

|  | CHF | EUR | GBP | JPY |
|---|---|---|---|---|
| Number of epochs | 2622 | 27370 | 14389 | 10181 |
| Number of patterns in training set: | 1200 | 1200 | 1200 | 1200 |
| Number of patterns in production set: | 103 | 103 | 103 | 103 |
| R squared: | 0,6529 | 0,6694 | 0,6714 | 0.6901 |
| r squared: | 0,6550 | 0,6697 | 0,6722 | 0.6917 |
| Mean squared error: | 0,003 | 0,001 | 0,003 | 0.002 |
| Mean absolute error: | 0,041 | 0,029 | 0,041 | 0.039 |
| Minimum absolute error: | 0 | 0 | 0,001 | 0 |
| Maximum absolute error: | 0,178 | 0,122 | 0,146 | 0.150 |
| Pearson's correlation coefficient: | 0,8093 | 0,8183 | 0,8199 | 0.8317 |
| Percent within 5%: | 37,879 | 52,427 | 35,960 | 39.798 |
| Percent within 5% to 10%: | 28,283 | 24,272 | 30,505 | 30.101 |
| Percent within 10% to 20%: | 26,010 | 23,301 | 26,263 | 24.848 |
| Percent within 20% to 30%: | 5,556 | 0 | 5,657 | 3.636 |
| Percent over 30%: | 2,273 | 0 | 1,616 | 1.616 |
| Probability of correct sign of increments forecasting: | 0,82 | 0,82 | 0,77 | 0.83 |

Table 3. Statistics of forecast quality for main currencies of Forex for the network with added technical indicators

|  | CHF | EUR | GBP | JPY |
|---|---|---|---|---|
| Number of epochs | 3921 | 149 | 2404 | 12581 |
| Number of patterns in training set: | 1600 | 1200 | 1600 | 1600 |
| Number of patterns in production set: | 100 | 100 | 100 | 100 |
| R squared: | 0.6836 | 0.4530 | 0.5436 | 0.6168 |
| r squared: | 0.6941 | 0.4584 | 0.5503 | 0.6212 |
| Mean squared error: | 0.002 | 0.002 | 0.003 | 0.002 |
| Mean absolute error: | 0.037 | 0.038 | 0.040 | 0.034 |
| Minimum absolute error: | 0 | 0.001 | 0.001 | 0 |
| Maximum absolute error: | 0.155 | 0.110 | 0.139 | 0.137 |
| Pearson's correlation coefficient: | 0.8332 | 0.6770 | 0.7418 | 0.7882 |
| Percent within 5%: | 45.299 | 50.000 | 35,960 | 39.655 |
| Percent within 5% to 10%: | 22.222 | 20.192 | 29.310 | 34.483 |
| Percent within 10% to 20%: | 26.496 | 21.154 | 25.000 | 22.414 |
| Percent within 20% to 30%: | 3.419 | 7.692 | 6.897 | 2.586 |
| Percent over 30%: | 2.564 | 0.962 | 0.862 | 0.862 |
| Probability of correct sign of increments forecasting: | 0,78 | 0,78 | 0,73 | 0.81 |

Table 4. Statistics of forecast quality for main currencies of Forex for the network with two added indicators and without moving average and returns.

|  | CHF | EUR | GBP | JPY |
|---|---|---|---|---|
| Number of epochs | 1918 | 114 | 6271 | 136 |
| Number of patterns in training set: | 1600 | 1200 | 1600 | 1600 |
| Number of patterns in production set: | 100 | 100 | 100 | 100 |
| R squared: | 0.5153 | 0.4032 | 0.4114 | 0.4693 |
| r squared: | 0.5332 | 0.4061 | 0.4139 | 0.4798 |
| Mean squared error: | 0.003 | 0.003 | 0.003 | 0.003 |
| Mean absolute error: | 0.045 | 0.039 | 0.047 | 0.041 |
| Minimum absolute error: | 0 | 0 | 0 | 0 |
| Maximum absolute error: | 0.168 | 0.130 | 0.142 | 0.142 |
| Pearson's correlation coefficient: | 0.7302 | 0.6373 | 0.6433 | 0.6927 |
| Percent within 5%: | 37.607 | 42.308 | 29.310 | 36.207 |
| Percent within 5% to 10%: | 23.077 | 24.038 | 33.621 | 31.897 |
| Percent within 10% to 20%: | 26.496 | 25.000 | 28.448 | 24.138 |
| Percent within 20% to 30%: | 7.692 | 7.692 | 5.172 | 7.759 |
| Percent over 30%: | 5.128 | 0.962 | 3.448 | 0 |
| Probability of correct sign of increments forecasting: | 0,74 | 0,75 | 0,73 | 0,80 |

Table 5. Statistics of forecast quality for main currencies of Forex for the network without raw data in input

|  | CHF | EUR | GBP | JPY |
|---|---|---|---|---|
| Number of epochs | 3794 | 149 | 1885 | 1997 |
| Number of patterns in training set: | 1600 | 1200 | 1600 | 1600 |
| Number of patterns in production set: | 100 | 100 | 100 | 100 |
| R squared: | 0.6513 | 0.4541 | 0.5124 | 0.6022 |
| r squared: | 0.6614 | 0.4601 | 0.5177 | 0.6051 |
| Mean squared error: | 0.002 | 0.002 | 0.003 | 0.002 |
| Mean absolute error: | 0.039 | 0.038 | 0.042 | 0.036 |
| Minimum absolute error: | 0 | 0 | 0 | 0 |
| Maximum absolute error: | 0.166 | 0.108 | 0.140 | 0.126 |
| Pearson's correlation coefficient: | 0.8132 | 0.6783 | 0.7195 | 0.7779 |
| Percent within 5%: | 39.316 | 48.077 | 37.931 | 41.379 |
| Percent within 5% to 10%: | 27.350 | 21.154 | 27.586 | 31.897 |
| Percent within 10% to 20%: | 25.641 | 23.077 | 26.724 | 24.138 |
| Percent within 20% to 30%: | 5.128 | 6.731 | 6.897 | 2.586 |
| Percent over 30%: | 2.564 | 0.962 | 0.862 | 0 |
| Probability of correct sign of increments forecasting: | 0,78 | 0,76 | 0,71 | 0,81 |

Table 6. Statistics of forecast quality for main currencies of Forex for the network with exponential moving average in input.

|  | CHF | EUR | GBP | JPY |
|---|---|---|---|---|
| Number of epochs | 4143 | 132 | 3518 | 117 |
| Number of patterns in training set: | 1600 | 1200 | 1600 | 1600 |
| Number of patterns in production set: | 100 | 100 | 100 | 100 |
| R squared: | 0.6825 | 0.4470 | 0.5106 | 0.6954 |
| r squared: | 0.6841 | 0.4384 | 0.5116 | 0.7024 |
| Mean squared error: | 0.002 | 0.002 | 0.003 | 0.001 |
| Mean absolute error: | 0.037 | 0.038 | 0.042 | 0.032 |
| Minimum absolute error: | 0.001 | 0.001 | 0 | 0 |
| Maximum absolute error: | 0.128 | 0.110 | 0.144 | 0.084 |
| Pearson's correlation coefficient: | 0.8271 | 0.6580 | 0.7152 | 0.8381 |
| Percent within 5%: | 40.171 | 50.000 | 35.345 | 40.517 |
| Percent within 5% to 10%: | 31.624 | 20.192 | 31.897 | 39.655 |
| Percent within 10% to 20%: | 22.222 | 21.154 | 25.000 | 18.103 |
| Percent within 20% to 30%: | 3.419 | 7.692 | 6.034 | 1.724 |
| Percent over 30%: | 2.564 | 0.962 | 1.724 | 0 |
| Probability of correct sign of increments forecasting: | 0,79 | 0,78 | 0,78 | 0,89 |

Table 7. Statistics of forecast quality for main currencies of Forex for the network with four values of returns in series

|  | CHF | EUR | GBP | JPY |
|---|---|---|---|---|
| Number of epochs | 2282 | 1764 | 3517 | 3513 |
| Number of patterns in training set: | 1600 | 1200 | 1600 | 1600 |
| Number of patterns in production set: | 100 | 100 | 100 | 100 |
| R squared: | 0.6369 | 0.5946 | 0.5103 | 0.5678 |
| r squared: | 0.6476 | 0.6030 | 0.5212 | 0.5706 |
| Mean squared error: | 0.003 | 0.002 | 0.003 | 0.002 |
| Mean absolute error: | 0.039 | 0.031 | 0.042 | 0.036 |
| Minimum absolute error: | 0.001 | 0.001 | 0.001 | 0 |
| Maximum absolute error: | 0.157 | 0.102 | 0.143 | 0.138 |
| Pearson's correlation coefficient: | 0.8048 | 0.7765 | 0.7220 | 0.7554 |
| Percent within 5%: | 43.363 | 45.545 | 39.823 | 46.018 |
| Percent within 5% to 10%: | 19.469 | 34.653 | 23.894 | 23.009 |
| Percent within 10% to 20%: | 30.973 | 14.851 | 29.204 | 28.319 |
| Percent within 20% to 30%: | 3.540 | 4.950 | 6.195 | 2.655 |
| Percent over 30%: | 2.655 | 0 | 0.885 | 0 |
| Probability of correct sign of increments forecasting: | 0,75 | 0,78 | 0,72 | 0,78 |